\documentclass{amsart}

\usepackage{amsfonts,amssymb}
\usepackage{cite}
\usepackage{graphicx,amsmath,latexsym}
\usepackage{epsf,color}
\usepackage{graphics}
\usepackage{url}

\usepackage{subfigure}
\newtheorem{theorem}{Theorem}[section]
\newtheorem{proposition}{Proposition}[section]

\def\bp{\noindent{\it Proof.}\ }
\def\ep{\hfill $\Box$}

\usepackage{bbm}

\def\E{\mathbb{E}}
\def\P{\mathbb{P}}
\def\R{\mathbb{R}}

\def\N{\mathbb{N}}

\def\Ncal{\mathcal{N}}

\newcommand{\indicator}[1]{\mathbbm{1}_{\{#1\}}}

\newcommand{\stochless}{\preceq}

\title[Load Balancing via Random Local Search in Closed and Open systems]{Load Balancing via Random Local Search in Closed and Open systems}

\author{A.\  Ganesh}
\address{Department of Mathematics, University of Bristol, UK}
\email{a.ganesh@bristol.ac.uk}

\author{S.\ Lilienthal}
\address{Stats Lab, Cambridge University, UK}
\email{s.lilienthal@statslab.cam.ac.uk}

\author{D.\ Manjunath}
\address{Department of EE, IIT Mumbai, India}
\email{dmanju@ee.iitb.ac.in}

\author{A.\ Prouti\`ere}
\address{Microsoft Research, Cambridge, UK}
\email{aproutie@microsoft.com}

\author{F.\ Simatos}
\address{INRIA Paris-Rocquencourt, Domaine de Voluceau, 78153 Le Chesnay, France}
\email{Florian.Simatos@inria.fr}


\begin{document}

\maketitle
\begin{abstract}
In this paper, we analyze the performance of random {\it load resampling 
and migration} strategies in parallel server systems. Clients initially 
attach to an arbitrary server, but may switch servers independently at 
random instants of time in an attempt to improve their service rate. This 
approach to load balancing contrasts with traditional approaches where 
clients make smart server selections upon arrival (e.g., 
Join-the-Shortest-Queue policy and variants thereof). Load resampling is 
particularly relevant in scenarios where clients cannot predict the load of 
a server before being actually attached to it. An important example is in 
wireless spectrum sharing where clients try to share a set of frequency 
bands in a distributed manner.

We first analyze the natural {\it Random Local Search (RLS)} strategy. 
Under this strategy, after sampling a new server randomly, clients only 
switch to it if their service rate is improved. In closed systems, where 
the client population is fixed, we derive tight estimates of the time it 
takes under RLS strategy to balance the load across servers. We then study 
open systems where clients arrive according to a random process and leave 
the system upon service completion. In this scenario, we analyze how client 
migrations within the system interact with the system dynamics induced by 
client arrivals and departures. We compare the load-aware RLS strategy to a 
load-oblivious strategy in which clients just randomly switch server 
without accounting for the server loads. Surprisingly, we show that both 
load-oblivious and load-aware strategies stabilize the system whenever this 
is at all possible. We further demonstrate, using large-system asymptotics, 
that the average client sojourn time under the load-oblivious strategy is 
not considerably reduced when\\ clients apply smarter load-aware strategies.
\end{abstract}

\section{Introduction}

Load balancing is a key component of today's communication networks and computer systems in which resources are distributed over a wide area or across a large number of systems and have to be shared by a large number of users. Load balancing enables efficient resource utilization and thereby tends to improve the quality of service perceived by users. Traditionally, load balancing has been achieved by applying smart routing policies: when a new demand arrives, it is routed towards a particular resource depending on the current loads of the various resources, see \cite{mor} and references therein. In contrast, we are interested in systems where a new task is initially assigned to a resource chosen at random irrespective of the current resource loads, but where tasks can be re-assigned, i.e., migrate from one resource to another. 

Our primary motivation stems from the increasing popularity of Dynamic Spectrum Access (DSA) techniques \cite{dyspan} as a potential mechanism for broadband access in future wireless systems. A common implementation platform for DSA is the use of reprogrammable Software-Defined-Radios (SDRs). These new radios are {\it frequency-agile} or {\it flexible}, and have the ability to rapidly jump from one frequency band to another in order to explore and exploit large parts of the spectrum. A central question in DSA is how multiple users may fairly and efficiently share spectrum in a distributed manner. Typically, the service rate of a user on a given frequency band is inversely proportional to the number of users transmitting on this band, i.e., to the load of the band. As new users entering the system have no way of determining the load on each frequency band, they have to initially select a band randomly. Should a user receive a quite poor quality of service on a given band, she may resample a new band at random and decide to switch to it. The overall system performance then strongly depends on the distributed resampling and switching strategy implemented by each user. 

Though our primary motivation is DSA, our methods and results could provide insight into a number of other applications. One such pertains to wireline networks, where there has recently been interest in multipath routing \cite{MassKeyTow}. Here, users may use several path to download files, and have to select the appropriate path or the set of paths. Another application is in transport networks, where one might wish to understand how Wardrop equilibria, which correspond to the equalization of journey times across alternative routes, are achieved or approximated by network users acting on limited information. Our results could also shed insight on how quickly such equilibria can be re-established following major disruptions or other changes to the network. Finally, note that distributed load resampling can also be thought of as a game between selfish users. In fact, it is an instance of a congestion game (see e.g. \cite{nisan}), and our results shed light on the time to reach a Nash equilibrium in such a game, but it also helps understanding the outcome of the game with a dynamic population of players.

We consider a generic system consisting of multiple servers (in DSA, frequency bands) employing the Processor Sharing (PS) service discipline, shared by clients who have to initially pick a server at random, and may later resample servers and migrate during their service. We restrict our attention to two natural distributed resampling and migration strategies, the  {\it Random Local Search (RLS)} and {\it Random Load-Oblivious (RLO)} strategies. When implementing the RLS algorithm, a user resamples a new randomly chosen server at the instants of a Poisson process, and migrates to this new server if its load is smaller than that of the initial server. In contrast, under the RLO algorithm, a user hops between servers according to a random Markovian jump process irrespective of the loads of the visited servers.  

We investigate both {\it closed} systems with fixed population of clients, and {\it open} systems with a population whose dynamics are governed by client arrivals and the completions of their services. In closed systems, we are interested in characterizing the time that it takes under the RLS algorithm to balance all server loads (note that here the RLO algorithm does not balance loads except in an average sense -- so we do not study this algorithm in closed systems). In open systems, users arrive at the various servers according to independent stochastic processes of fixed intensities, and leave  upon service completion. In this scenario, client migrations within the system interact in a complicated manner with the system dynamics induced by client arrivals and departures. We aim at characterizing system stability under the RLS and RLO strategies, as well as at deriving estimates of user sojourn times. Our contributions are as follows:
\begin{itemize}
\item {\it Closed systems.} We show that, starting from an arbitrary allocation of users to servers, the time $\tau$ it takes to achieve perfect balance of server loads scales at most as $\log(m)\big({m^2\over n}+\log(m)\big)$, where $m$ and $n$ denotes the number of servers and users, respectively. This considerably improves over the existing bounds that stated that $\tau$ scales at most as $m^2$ (see e.g. \cite{Goldberg04}). We also investigate the time $\tau_\epsilon$ to reach an approximate $\epsilon$-balance (a system reaches an approximate $\epsilon$-balance if there exists $p$ such that the number of users associated to any server lies between $(1-\epsilon)p$ and $(1+\epsilon)p$). Achieving such balance is much faster than reaching a perfect balance, and we show that $\tau_\epsilon$ scales at most as $\log(m)/\epsilon$.  
\item {\it Open systems.} We demonstrate that both RLS and RLO strategies achieve the largest stability region possible, i.e., that the system is stable under these two algorithms provided that $\sum_{i=1}^m\lambda_i < \sum_{i=1}^m\mu_i$, where $\lambda_i$ denotes the initial user arrival rate at server $i$ and $\mu_i$ is the service rate of this server. The result is not surprising for RLS, but less intuitive for RLO since, under this algorithm, users take no account of server loads when migrating. For both RLS and RLO strategies, we derive approximate estimates of the average user sojourn time using large-system asymptotics. The estimates are shown to be exact when the number of servers grows large, but turn out to be quite accurate for systems of limited sizes as well. Our first numerical results suggest that again, surprisingly, the average client sojourn time under the load-oblivious RLO strategy is not considerably reduced when clients apply smarter load-aware RLS strategy. To our knowledge, this paper is the first to analyze the performance of RLS and RLO algorithms in open systems.
\end{itemize}       
The paper is organized as follows. In the next section, we describe our model and notation. Sections 3 and 4 are devoted to the analysis of closed and open systems, respectively. We give the related work in Section 5, and provide concluding remarks in Section 6.

\section{Model description and notation}

We consider a set of $m$ Processor Sharing servers of respective capacities $\mu_1,\ldots,\mu_m$. The system is {\it homogeneous} if $\mu_i=1$ for all $i=1,\ldots,m$. The system state at time $t$ is represented by the number of clients associated to each server, $N(t)=(N_1(t),\ldots,N_m(t))$. The service rate of a client associated to server $i$ at time $t$ is then $\mu_i/N_i(t)$. Clients independently resample and switch servers to selfishly improve their service rate. They have a myopic view of the system in the sense that they are aware of their current service rates, but do not know the service rate they would achieve at other servers. Given this myopic view, it is natural to consider and analyze the two following random distributed resampling and migration algorithms:
\begin{itemize}
\item {\it Random Local Search (RLS) algorithm.} At the instants of a Poisson process of intensity $\beta >0$, a client picks a new server uniformly at random and migrates to it if and only if this would increase her service rate. In other words, if at time $t$, a client associated to server  $i$ picks server $j$, she migrates to $j$ if and only if $\mu_j/(N_j(t)+1) > \mu_i/N_i(t)$.
\item {\it Random Load-Oblivious (RLO) algorithm.} After arriving in the system, each client visits successive servers according to a continuous-time random walk with transition matrix $Q=\{ q_{ij}, i,j=1,\ldots,m \}$. The random walks are independent across clients, and irreducible. We denote by $\pi$ the stationary distribution of this random walk. Note that as a consequence of irreducibility, clients visit all servers eventually, i.e., $\pi_i>0$ for all $i=1,\ldots, m$. 
\end{itemize}
Note that under the RLO algorithm, clients do not take loads into account when switching servers. In particular they may move to a server with a higher load. An example of such a resampling strategy is as follows. Each client has a Poisson clock of rate $\beta>0$ and, when her clock ticks, she picks a new server uniformly at random and moves there irrespective of its load. 

We analyze the performance of distributed resampling and migration strategies in closed and open systems. In closed systems, the total population of clients is fixed, equal to $n$. For such systems, we investigate the time it takes under the RLS algorithm to balance clients across servers, starting from any arbitrary system state. In open systems, exogenous clients associate to server $i$ according to a Poisson process of intensity $\lambda_i$ (the arrival processes are independent across servers). Client service requirements are i.i.d. exponentially distributed with unit mean. Under RLS and RLO algorithms, $(N(t),t\ge 0)$ is a Markov process. In open systems, we are interested in characterizing the {\it stability region} of RLS and RLO strategies, defined as the set of arrival rates $\lambda=(\lambda_1,\ldots,\lambda_m)$ such that the system is stable, i.e., such that $(N(t),t\ge 0)$ is positive recurrent. We also aim at estimating the average client sojourn time.

\section{Closed systems}

In this section, we analyze the performance of the RLS resampling
strategy in a closed homogeneous system, and obtain tight bounds on the expected time to balance the server loads. 

Recall that there are $n$ clients distributed among $m$ servers. 
Let $n= qm + r,$ $0 \leq r \leq m-1.$ We now define the following:
\begin{itemize}
\item The state $N(t)=(N_1(t),\ldots,N_m(t))$ is \emph{balanced} 
if $|N_i(t)-N_j(t)| \le 1$ for $1 \leq i < j \leq m.$ The time 
to balance, $\tau,$ is defined as
  \begin{displaymath}
    \tau := \inf \{ t>0: \mbox{ $N(t)$ is balanced} \}.
  \end{displaymath}
\item The state $N(t)$ is \emph{$\epsilon$-balanced} if
$(1-\epsilon)p\le N_i(t) \le (1+\epsilon)p$ for all $i=1,\ldots,m$,
where $p=n/m$. The time, $\tau_{\epsilon}$, to $\epsilon$-balance 
is defined as
  \begin{displaymath}
    \tau_{\epsilon} := \inf \{ t>0: \mbox{ $N(t)$ is $\epsilon$-balanced} \}.
  \end{displaymath}
\end{itemize}

Let $f,g:\N \to \R_+$. We say $f(k)=O(g(k))$ if there exist $k_0\in \N$ and $c\in \R_+$ such that $f(k) \le cg(k)$ for all $k\ge k_0$. Similarly, for $f,g:\N^2 \to \R_+$, we say $f(k,l)=O(g(k,l))$ if there exist $k_0,l_0\in \N$ and $c\in \R_+$ such that $f(k,l) \le cg(k,l)$ for all $k\ge k_0$ and $l\ge l_0$.

\subsection{Time to reach balance}

We now characterize the time required by the RLS algorithm to reach perfect balance and $\epsilon$-balance. 

\begin{theorem}
  The expected time, $\E[\tau]$, for randomized local search to achieve balance is 
  $O(\log(m) (\frac{m^2}{n}+\log(m)))$.
  \label{thm:mean-to-balance}
\end{theorem}

\begin{theorem}
  The expected time, $\E[\tau_{\epsilon}]$ for randomized local search to achieve $\epsilon$-balance is $O(\log(m)/\epsilon).$
  \label{thm:epsilon-balance}
\end{theorem}

\noindent \emph{Remarks}
\begin{enumerate}
\item It is easy to see, applying Markov's inequality, that the same upper bounds
on the time to balance hold in probability as in expectation.
\item We now compare our bounds on $\tau$ with that from
  \cite{Goldberg04}. From Theorem~$2.7$ of \cite{Goldberg04}, the
  expected number of attempted moves before reaching balance is $O(m^2n)$.
  Since move attempts (resampling) occur at rate $n,$ this gives us a time 
  complexity of $O(m^2)$. Our bound in Theorem \ref{thm:mean-to-balance} is
  much tighter. 
\item Our bound is close to the best possible. To see this, suppose
  $m$ divides $n$ exactly. At some stage, the algorithm will reach an
  allocation in which one server has $n/m+1$ clients, one other server
  has $n/m-1$ clients and all others have exactly $n/m$ clients. Each
  of the $n/m+1$ clients at the overloaded server attempts to move at
  rate 1, and each move attempt is successful with probability
  $1/m$. Hence, the mean time for just the final move is $m^2/(m+n)
  \ge m^2/(2n)$. Our bound is only a $\log m$ factor higher than the
  time for the last move.

  Alternatively, consider the situation when $m^2=o(n)$ and all $n$
  clients are initially at the same server. Then, at least $n-\lceil
  n/m \rceil$ clients need to move out of this server to reach balance.
  When there are $k$ clients at the server, the expected time to the
  next move is at least $1/k$ (possibly more, as the move attempt may
  not be successful). Hence, the expected time to reach balance is at
  least
  \begin{displaymath}
    \sum_{k=\lceil n/m \rceil+1}^n \frac{1}{k} \ge \int_{n/m}^n
    \frac{1}{x}dx = \log m.      
  \end{displaymath}
Again our bound is only a $\log m$ factor higher than the above lower bound on the time to reach balance.
\end{enumerate}

\subsection{Proofs}

Without loss of generality, we take the rate $\beta$ of the independent 
Poisson clocks at each client to be unity. A client at server $i$ whose 
clock has ticked at time $t$ attempts to move by sampling a server uniformly 
at random from all $m$ servers. It moves to the sampled server, say $j,$ 
if and only if $N_i(t) - N_j(t) > 1.$ Clearly,  $N(t)$ evolves as a continuous 
time Markov chain. 

\subsubsection{Proof of Theorem~\ref{thm:mean-to-balance}}

Define $ V(t) := \max_{1 \leq j \leq m} \ N_j(t),$ i.e., $V(t)$ is the  maximum number of clients associated with any server at time $t.$ Define $C_v(t)$ to be the number of servers with exactly $v$ clients, $B_v(t)$ to be the number with exactly $v-1$ clients and $A_v(t)$ to be the number with strictly less than $v-1$ clients, all at time $t$.

The idea of the proof is as follows. The evolution of $N(t)$ towards
balance is divided into phases. If $V(t) = v,$ then $N(t)$ is said to
be in phase $v.$ Thus, $C_v(t)$ is the number of maximally loaded
servers in phase $v.$ Since a client never moves to a server that has
more clients than its current server, $V(t)$ is monotone decreasing and,
in each phase, $C_v(t)$ is also monotone decreasing. Phase $v$ ends
when $C_v(t)=0.$ Let $\tau_v$ denote the (random) length of phase $v.$
Each phase can be further divided into sub-phases, say $(v,c),$ when $C_v(t)=c.$
Let $\tau_{v,c}$ denote the random length of time that it takes for
$C_v(t)$ to decrease from $c$ to $c-1$. Observe that $\tau_v = \sum_c
\tau_{v,c}$ and $\tau = \sum_{v} \tau_v.$ When $N(t)$ is balanced,
$V(t) = \lceil \frac{n}{m} \rceil$, $C_{\lceil \frac{n}{m} \rceil}(t) = r$ if $r>0$ and $C_{\lceil \frac{n}{m} \rceil}(t) = m$ otherwise. This gives us the
maximum range for $v.$ The number of sub-phases in each phase is also
similarly bounded. The theorem is proved by bounding the expected
times of each of the sub-phases and phases.

\begin{proof}
  In phase $v,$ observe that 
  \[
    vC_v(t)+(v-1)B_v(t) \le n \ \text{ and }\ m - B_v(t) - C_v(t) =  A_v(t).
  \]
  Further, for $\lceil n/m \rceil \leq v \leq \lceil n/(m-1) \rceil$,
  $n/(v-1) \in (m-1,m]$, but if $N(t)$ is not balanced, then there has
  to be at least one server with $v-2$ or fewer clients (i.e., $A_v(t) \geq
  1$). Hence
  \begin{equation}
    \label{lbd_light_queues1}
    A_v(t) \ge \max \Bigl\{ m - \frac{n}{v-1}, 1 \Bigr\}.
  \end{equation}
  Each of the $vC_v(t)$ clients at one of the maximally loaded servers
  samples one of the $m$ servers at random at unit rate. If the sampled
  server happens to be one of the $A_v(t)$ servers with $v-2$ or fewer
  clients, then the client moves to the sampled server and $C_v(t)$
  decreases by 1. This event has probability $A_v(t)/m$. Hence,
  $C_v(t)$ decreases by 1 at a rate no smaller than $vC_v(t) A_v(t)/m$
  and from (\ref{lbd_light_queues1}), we obtain that $\tau_{v,c} \stochless {\tilde \tau}_{v,c} \sim Exp(\lambda_{v,c})$ where
  \begin{equation} 
    \lambda_{v,c} := v C_v(t) \frac{A_v(t)}{m} = \ \Bigl( vc \Bigl( \max \Bigl\{ 1-\frac{n}{m(v-1)},
    \frac{1}{m} \Bigr\} \Bigr) \Bigr). \label{tauvc_bd1} 
  \end{equation}
  Here we write $X \stochless Y$ to mean that $X$ is stochastically
  dominated by $Y,$ (i.e., for all $t$, $\P[X>t] \le \P[Y>t]$), $X\sim Y$ to mean that they have the same
  distribution and $Exp(x)$ to denote an exponentially distributed
  random variable with rate $x.$ In particular,
  \begin{equation}
    \E[\tau_{v,c}] \le \E[{\tilde \tau}_{v,c}] \le \frac{1}{vc} \min
    \Bigl\{ \frac{ m(v-1) }{ [m(v-1)-n]^+ }, m
    \Bigr\}, \label{etauvc_bd} 
  \end{equation}
  where $x^+$ denotes $\max \{x,0\}$.  

  At any time $t$, $C_v(t)$ is bounded above by $\lfloor n/v
  \rfloor$, since there cannot be more than this many servers with $v$
  clients. Since phase $v$ ends when $C_v=0,$ we have $\E[\tau_v] \ \le
  \ \sum_{c=1}^{\lfloor n/v \rfloor} \tau_{v,c},$ and we obtain
  \begin{equation}
    \label{etauv_bd1}
    \E[\tau_v] \ \le \ \min \Bigl\{ \frac{ m(v-1) }{ [m(v-1)-n]^+ }, m
    \Bigr\} \frac{1}{v} \sum_{c=1}^{\lfloor n/v \rfloor} \frac{1}{c}. 
  \end{equation}

Finally, $\tau$, the time it takes to reach perfect balance, satisfies
  \begin{equation} 
    \label{tau_bd1} 
    \tau \le \sum_{v=\lceil n/m
      \rceil+1}^n \tau_v + \sum_{c=r+1}^m \tau_{\lceil n/m \rceil,c}.
  \end{equation}
  Now, we have by (\ref{etauvc_bd}) that,
  \begin{align}
    \sum_{c=r+1}^m \E[\tau_{\lceil n/m \rceil,c}] &\le \frac{m}{\lceil
      n/m \rceil} \sum_{c=r+1}^m \frac{1}{c} \nonumber \\
    & \le \frac{m^2}{n} \Bigl( 1+\int_1^m \frac{1}{x}dx \Bigr) = \frac{m^2}{n} (1+\log m). \label{last_term_bd}
  \end{align}
  For all $v \ge \lceil \frac{n}{m} \rceil,$ we also readily see that,
  \[
    \sum_{c=1}^{\lfloor n/v \rfloor} \frac{1}{c} \le 1+\int_1^{n/v} \frac{1}{x} dx \le 1+\log \frac{n}{v} \le 1+\log m . 
  \]
  Hence, from (\ref{etauv_bd1}), (\ref{tau_bd1}) and
  (\ref{last_term_bd}), we obtain
  \begin{equation}
    \frac{\E[\tau]}{1+\log m} \le \frac{m^2}{n} + \sum_{v=\lceil n/m \rceil+1}^{\lceil n/(m-1) \rceil} \frac{m}{v} + \sum_{v=\lceil n/(m-1) \rceil+1}^n \frac{m}{m(v-1)-n}. \label{etau_bd2} 
  \end{equation}
  The number of terms in the first sum above is at most $\max\{1,
  \frac{n}{m(m-1)} \}$. Each summand is no more than $m^2/n$. Hence,
  the first sum is bounded above by $\max \{ \frac{m^2}{n}, 2 \}$. The
  second sum is bounded above by
  \[
    \frac{m}{\frac{mn}{m-1}-n} + \int_{\frac{n}{m-1}+1}^{n}
    \frac{m}{m(x-1)-n} dx = \frac{m(m-1)}{n} + \log \frac{ m(n-1)-n }{ 
      \frac{mn}{m-1}-n }. 
  \]
  Substituting these expressions in (\ref{etau_bd2}) and simplifying,
  we get
  \begin{align*}
    \E[\tau] &\le (1+\log m) \Bigl( \max \{ \frac{m^2}{n}, 2 \} +
    \frac{m^2}{n} + \log(m^2) + \frac{m^2}{n} \Bigr) \\
    &\le 3(1+\log m) \Bigl( \frac{m^2}{n}+\log m + 1 \Bigr).
  \end{align*}
  This completes the proof. 
\end{proof}

\subsubsection{Proof of Theorem~\ref{thm:epsilon-balance}}

We need the following definitions.
\begin{itemize}
\item Let $p=n/m$. Server $i$ is $\epsilon$-balanced at time $t$ if $(1-\epsilon)p
  \leq N_i(t) \leq (1 + \epsilon) p,$ underloaded if $N_i(t) <
  (1-\epsilon)p$ and overloaded if $N_i(t) > (1+\epsilon)p.$ $M_C(t),$
  $M_U(t)$ and $M_O(t)$ denote the number of $\epsilon$-balanced,
  underloaded and overloaded servers, respectively.
\item The underflow from server $i$ is defined to be
  \begin{displaymath}
    u_i(t) = 
    \begin{cases}
      0  & \mbox{if $N_i(t) \geq p$} \\
      p - N_i(t) & \mbox{otherwise.}
    \end{cases}
  \end{displaymath}
  Also, let $U(t) := \sum_{i=1}^m u_i(t).$ Similarly, define the
  overflow from server $i$ as
  \begin{displaymath}
    o_i(t) = 
    \begin{cases}
      0  & \mbox{if $N_i(t) \leq p$} \\
      N_i(t) - p & \mbox{otherwise,}
    \end{cases}
  \end{displaymath}
  and $O(t) := \sum_{i=1}^m o_i(t).$  
\end{itemize}

\begin{proof}
  Let $N_O(t)$ be the number of `overflowing' clients defined as
  \begin{displaymath}
    N_O(t) := \sum_{i\in {\mathcal{M}}_O(t)} \left(N_i(t) -
      (1+\epsilon)p \right), 
  \end{displaymath}
  where ${\mathcal{M}}_O(t)$ is the set of overloaded servers at time $t$.
  We can write
  \begin{align*}
    U(t) & \leq (M_U(t) \times p) + M_C(t) \times (\epsilon p),\\
    O(t) & \geq N_O(t) + (m - M_U(t) - M_c(t)) \times (\epsilon p).
  \end{align*}
  Since $O(t) = U(t),$ we obtain
  \[
    p \ M_U(t)  + (\epsilon p) M_C(t) \geq \ N_O(t) + (m - M_u(t) - M_C(t)) (\epsilon p) ,
  \]
  which yields
  \begin{align*}
    N_O(t) & \leq p (M_U(t) + M_C(t))(1+\epsilon) - \epsilon m ) \\ 
    M_C(t) + M_U(t) & \geq \frac{N_O(t)}{(1+\epsilon)p} + \frac{\epsilon}{1+\epsilon} m \geq \max \left\{ \frac{N_O(t)}{(1+\epsilon)p}, \frac{\epsilon}{1+\epsilon} m \right\} . 
  \end{align*}
  Now consider a client that is attempting to move at time $t.$ We say
  that this attempt results in a good move if the
  attempt results in a migration that reduces $N_O(t).$ Let $G$ denote the event corresponding to a good move. When the state of
  the system is $(N_O,M_C,M_U)$, the probability of a good move is
  \begin{displaymath}
    \P\left(G\right) \ \geq \ \frac{N_O + (1+\epsilon)p}{n} \ \frac{M_C
      + M_U }{m}
  \end{displaymath}
  and the number of attempts between successive good moves is
  geometric with mean at most $\frac{m n}{ (N_O + p)M_U}.$

  Let $K_G$ denote the number of attempts before a good move occurs
  from the state $(N_O,M_C,M_U).$ The expected number of attempts
  before a good move reduces $N_O$ satisfies
  \begin{displaymath}
    \E[K_G] \ \leq \ \frac{m n}{(N_O + (1+\epsilon)p)(M_C+ M_U)}.
  \end{displaymath}

  Let $K_{\epsilon}$ denote the number of attempts to achieve
  $\epsilon$-balance. In the worst case, $N_O(t)$ starts at $(m-1)p$
  and ends at $1.$ We can then bound $\E[K_{\epsilon}]$ as follows.
  \begin{align*}
    \E[K_\epsilon] & \leq \sum_{i=1}^{(m-1)p} \frac{m n}{( (1+\epsilon)p +  i) ( \max\{\frac{i}{(1+\epsilon)p},
      \frac{\epsilon}{1+\epsilon}m \} )} \\
    & = \sum_{i=1}^{\epsilon n} \frac{m n}{( (1+\epsilon) p + 
      i) \left(\frac{\epsilon}{1+\epsilon}m \right) } + \ \sum_{i=\epsilon n +1}^{(m-1)p} \frac{m n}{(
      (1+\epsilon) p + i) \left( \frac{i}{(1+\epsilon)p}\right)} \\ 
    & = \frac{1 + \epsilon}{\epsilon}  n \sum_{i=1}^{\epsilon n} \frac{1}{( (1+\epsilon) p +  i) } + m n \sum_{i=\epsilon n +1}^{(m-1)p} \frac{1}{i} -
    \frac{1}{(1+\epsilon) p +  i)} \\ 
    & \leq \frac{1 + \epsilon}{\epsilon} n \ \log \left( 
      \frac{(1+\epsilon)p + \epsilon n}{ (1+\epsilon)p } \right) + m n \log \left(\frac{(m-1)p}{\epsilon n} 
      \frac{(1+\epsilon)p + \epsilon n}{(1+\epsilon)p + (m-1)p}
    \right)\\
    & \leq \frac{n}{\epsilon}  \log (1+\epsilon m) + m n 
    \left(- \frac{1}{m} + \frac{1 + \epsilon}{\epsilon m} -
      \frac{\epsilon}{m} \right) \\
    & \leq \frac{n}{\epsilon}  \log (m) . 
  \end{align*} 
  Since each client is sampling at unit rate, the total sampling rate
  is $n$ and the average time to reach $\epsilon$-balance, $\E[\tau_{\epsilon}]$
  is $\E[K_{\epsilon}] / n.$ Thus $\E[\tau_{\epsilon}] = O((\log
  m)/\epsilon).$
\end{proof}

\section{Open systems}

In open systems, we are interested in quantifying classical queueing performance metrics, such as the stability region and the mean client sojourn time. We first investigate the stability region achieved under RLO and RLS algorithms. Both algorithms are shown to stabilize the system whenever this is at all possible, which for load-oblivious RLO algorithm may be surprising. Then, we try to obtain more detailed estimates of the system performance. As it turns out, the system equilibrium distribution is difficult, if not impossible, to derive, and we rely on large-system asymptotics to provide insights into the way the system behaves.

\subsection{Stability}

In the following, we denote $\lambda=(\lambda_1,\ldots,\lambda_m)$ and $\mu=(\mu_1,\ldots,\mu_m)$ the vectors representing the arrival and departure rates at the various servers. $\lVert \cdot\rVert$ denotes the $L_1$-norm on $\mathbb{R}^m$. We first provide an upper bound on the maximum stability region defined as the set of $\lambda$ such that there may exist a resampling and migration strategy stabilizing the system. This set is obtained by assuming that all servers' resources are pooled. 

\begin{proposition} Assume that $\lambda$ is such that $\sum_i\lambda_i > \sum_i\mu_i$. Then there is no resampling and migration strategy stabilizing the system.
\end{proposition}

\bp The proof is straightforward. Remark that for any resampling and migration strategy, the total service rate is less than $\sum_i\mu_i$. Then if $\sum_i\lambda_i > \sum_i\mu_i$, the average number of clients in the system grows at a rate greater than $\sum_i\lambda_i - \sum_i\mu_i > 0$. The system is then unstable. \ep

The two following theorems state that both RLO and RLS strategies achieve maximum stability.

\begin{theorem}\label{th:stab1}
Assume that $\sum_i\lambda_i < \sum_i\mu_i$. Then the system is stable under RLO algorithm.
\end{theorem}

\begin{theorem}\label{th:stab2}
Assume that $\sum_i\lambda_i < \sum_i\mu_i$. Then the system is stable under RLS algorithm.
\end{theorem}

A result somehow similar to that of Theorem~\ref{th:stab1} was first stated in~\cite{Borst06:0} using heuristic fluid limits arguments. Fluid limits are powerful techniques to study ergodicity of Markov processes \cite{Dai95}. They comprise the study of the system behavior in the following limiting regime: the initial condition is scaled up by a multiplicative factor $k$, time is accelerated by the same factor, and $k$ tends to $\infty$. Often the system becomes tractable in this regime and even deterministic. If the system in the fluid regime reaches 0 in a finite time, then the process is ergodic. In the fluid regime, clients stay for very long periods of time in our system, and since, under RLO algorithm, the client random walks are ergodic, the probability that a given client is associated to server $i$ {\it should} be proportional to $\pi_i$ (the equilibrium distribution of the random walk). In such case, when the client population is large (as in the fluid regime), all servers should be occupied and active, ensuring that the system empties in finite time. This is the argument used in \cite{Borst06:0}, but not justified. The problem arises because the client migration process actually interacts with arrivals and departures. Handling this interaction turns out to be extremely difficult. Recently however, in \cite{Simatos09:0}, the authors were able to formally derive the system fluid limits, and analyze its stability under very specific assumptions on the client random walk (its transition matrix $Q$ has to be diagonalizable). Their proof is quite intricate. In the following, we prove Theorem~\ref{th:stab1} without the use of fluid limits, and for {\it any} random walk. Our proof is much more direct than that in~\cite{Simatos09:0}, and hence is amenable to deal with more general cases and possible extensions. For the proof of Theorem~\ref{th:stab2}, we use a rather classic method, i.e., we exhibit a simple Lyapunov function.

\subsubsection{Proof of Theorem \ref{th:stab1}}

Recall that by definition, under RLO strategy, the process $(N(t),t\ge 0)$ is the Markov process with the following non-zero transition rates for $1 \leq i \neq j \leq m$:
\begin{equation} \label{eq:rates}
\left \{ \begin{array}{l}
	\Omega(n, n+e_i) = \lambda_i,\\
	\Omega(n, n-e_i+e_j) = n_i q_{ij},\\
	\Omega(n, n-e_i) = \mu_i \indicator{n_i > 0},
\end{array} \right .
\end{equation}
where $n = (n_1, \ldots, n_m) \in \N^m$ and $e_i$ is the $m$-th dimensional vector with every coordinate equal to $0$, except for the $i$th one equal to $1$. The matrix $Q = (q_{ij})$ describes the migration of clients, and it is only assumed to possess a unique stationary distribution $\pi = (\pi_i)$ such that $\pi_i > 0$ for each $i = 1, \ldots, m$. The aim of the analysis is to use the following result, known as Foster's criterion~\cite{Robert03}.\\
(Foster's criterion) {\it If there exist $K$ and $t \geq 0$ such that
	\begin{equation} \label{eq:foster}
		\sup_{n \in \N^m: \lVert n \rVert \geq K} \E_n(\lVert N(t) \rVert - \lVert n \rVert) < 0,
	\end{equation}
	where $\E_n(\cdot)=\E(\cdot\vert N(0)=n)$, then $(N(t), t\ge 0)$ is ergodic.
}

Kolmogorov's equation is the first step that leads to~\eqref{eq:foster}: for any $t \geq 0$, the drift $\E_n(\lVert N(t) \rVert - \lVert n \rVert)$ is given by
\[
\E_n(\lVert N(t) \rVert - \lVert n \rVert) = \lVert \lambda \rVert t - \int_0^t \E_n \left( \sum_{i=1}^m \mu_i \indicator{N_i(u) > 0} \right) \, du.
\]
This gives the following inequality, which is the basis of our drift analysis:
\begin{equation} \label{eq:drift}
\E_n(\lVert N(t) \rVert - \lVert n \rVert) \leq \lVert \lambda \rVert t - \lVert \mu \rVert \int_0^t \P_n \left( N(u) > 0 \right) \, du,
\end{equation}
where $\P_n [\cdot] =\P[ \cdot \vert N(0)=n]$, and for $x \in \N^n$, $x > 0$ is to be understood coordinatewise, i.e., $x_i > 0$ for each $i = 1, \ldots, m$.

The idea of the proof of~\eqref{eq:foster} is that when the system starts with many clients, then the number of arrivals and departures is negligible on the time interval $[0,t]$ and the system behaves like the closed one. For a closed system, it is not difficult to show, using the fact that $Q$ has an invariant measure, that $\P(N(u) > 0)$ for $u > 0$ is arbitrarily close to~$1$ as the number of clients in the system increases. In view of~\eqref{eq:drift} this gives a negative drift when $\sum_i \lambda_i < \sum_i \mu_i$.

The following coupling initially proposed and formally justified in \cite{Simatos09:0}  is key to relate the open and closed systems. For $n \in \N^m$ and $\ell, \rho \in \R_+^m$, denote by $N^n_{\ell, \rho}$ the process under RLO strategy starting in the initial state $n$, with arrival rate~$\ell_i$ at server~$i$ with capacity $\rho_i$. Then $(N^n_{\ell, \rho}(t))$ is the Markov process with $N^n_{\ell, \rho}(0) = n$, and with non-zero transition rates given by~\eqref{eq:rates} with $\ell_i$ instead of $\lambda_i$ and $\rho_i$ instead of $\mu_i$. Then the processes $N^n_{\ell,0}$ and $N^0_{\rho, 0}$ can be coupled in such a way that for some process $Z(t) \geq 0$,
\[ N^n_{\ell, \rho}(t) = N^n_{\ell,0}(t) - N^0_{\rho, 0}(t) + Z(t), \ t \geq 0. \]
Moreover, the processes $\lVert N_{\rho, 0}^0 \rVert$ and $N_{\ell, 0}^n$ are independent, and $\lVert N_{\rho, 0}^0 \rVert$ is a Poisson process with parameter~$\lVert \rho \rVert$. Essentially, this coupling realizes the process $N^n_{\ell,\rho}$ with arrivals and departures as the difference between two processes without departures. This coupling can be constructed as follows: consider a particle system with three kinds of particles, colored blue, red and green. All the particles in the system are performing independent continuous-time random walks, going from $i$ to $j$ at rate $q_{ij}$, and the system starts with only blue particles.

Consider two independent Poisson processes $\Ncal_\ell$ and $\Ncal_\rho$ with respective parameters $\lVert \ell \rVert$ and $\lVert \rho \rVert$: at times of $\Ncal_\ell$, add a new blue particle at server $i$ with probability $\ell_i / \lVert \ell \rVert$. At times of $\Ncal_\rho$, consider server $i$ with probability $\rho_i / \lVert \rho \rVert$: if there is a blue particle, choose one at random and turn it into a red one. If there is no blue particle, add a green particle.

If $B_i(t), R_i(t)$ and $G_i(t)$ are respectively the number of blue, red and green particles at server $i$ at time $t$, then it is easy to see that:
\begin{itemize}
	\item $B$ is distributed like $N_{\ell, \rho}^n$,
	\item $B + R$ is distributed like $N_{\ell, 0}^n$,
	\item $R + G$ is distributed like $N_{\rho, 0}^0$ and $\lVert B + G \rVert = \Ncal_\rho$ is independent of $B+R$.
\end{itemize}
This proves the coupling with $Z(t)=G(t)$. The process $N^n_{\ell, 0}$ can be seen as the superposition of the initial particles with the particles arriving at rate $\lVert \ell \rVert$, hence the additional coupling $N^n_{\ell, 0} = N^n_{0,0} + N^0_{\ell,0}$ holds, and finally, $N_{\ell, \rho}^n$ can be written
\[ N^n_{\ell, \rho}(t) = N^n_{0,0}(t) + N^0_{\ell,0}(t) - N^0_{\rho, 0}(t) + Z(t), \ t \geq 0, \]
with $N^n_{0,0}$ and $\lVert N^0_{\rho, 0} \rVert$ independent, and $Z(t) \geq 0$. Starting from~\eqref{eq:drift}, we now turn our attention to proving the existence of constants $K$ and $t$ which satisfy~\eqref{eq:foster}. We have, using the coupling's notation, $\P_n(N(u) > 0) = \P(N_{\lambda, \mu}^n(u) > 0)$ and hence, for any $0 \leq u \leq t$ and $n \in \N^m$,
\begin{align*}
	\P_n(N(u) > 0) & = \P(N_{0,0}^n(u) + N_{\lambda, 0}^0(u) + Z(u) > N_{\mu, 0}^0(u))\\
	& \geq \P(N_{0, 0}^n(u) > \lVert N_{\mu, 0}^0(u) \rVert).
\end{align*}
Since the process $N_{0,0}^n$ is independent of the random variable $\lVert N_{\mu,0}^0(t) \rVert$, we can work conditionally on the value of $\lVert N_{\mu, 0}^n(t) \rVert$ and study the quantity $\P(N_{0, 0}^n(u) > M)$. Thus we only need to consider the closed process $N_{0,0}^n$ henceforth, and so we simplify the notation and note $N_{0,0}^n = N^n$. Markov's inequality gives
\begin{multline*}
	\P(\exists i \in \{1, \ldots, m\}: N_i^n(u) \leq M) = 1 - \P(N^n(u) > M)\\
	\leq \sum_{i=1}^m \P(N_i^n(u) \leq M) \leq e^{M} \sum_{i=1}^m \E \left( e^{-N_i^n(u)} \right).
\end{multline*}
For any $i \in \{ 1, \ldots, m \}$,
\[ \E \left( e^{-N_i^n(u)} \right) = \prod_{j=1}^m \left[ \E_j \left( e^{-\indicator{\xi(u) = i}} \right) \right]^{n_j} \]
where $\xi$ under $\P_j$ is a continuous-time Markov chain with transition rates $Q = (q_{ij})$, and which starts at $\xi(0) = j$. If $p(j,i,u) = \P_j(\xi(u) = i)$, one gets for $u \geq t_0 > 0$ and $n \in \N^m$ with $\lVert n \rVert \geq K$
\begin{align*}
	\E \left( e^{-N_i^n(u)} \right) & = e^{\sum_{j=1}^m n_j \log \left(1 - (1-1/e) p(j,i,u) \right)}\\
	& \leq e^{-\lVert n \rVert (1-1/e) p(t_0)} \leq e^{-K (1-1/e) p(t_0)}
\end{align*}
with $p(t_0) = \inf_{u \geq t_0} \min_{1 \leq i,j \leq m} p(j,i,u)$. Note that since, for any $1 \leq i, j \leq m$, $p(j,i,u) > 0$ for any $u > 0$ and $p(j,i,u) \to \pi_i > 0$ as $u \to +\infty$, one has that $p(t_0) > 0$. Therefore, for $u \geq t_0$ and $n$ with $\lVert n \rVert \leq K$, integrating on the law of $\lVert N_{\mu,0}^0(t) \rVert$ gives
\[
	\P(N^n(u) > \lVert N_{\mu,0}^0(t) \rVert) \geq 1 - \varepsilon(t, K, t_0)
\]
with $\varepsilon(t, K, t_0) = m e^{ \lVert \mu \rVert t (e-1) - K (1-1/e) p(t_0) }$. In particular, for $t \geq t_0$,
\[
\sup_{n \in \N^m: \lVert n \rVert \geq K} \E_n \left( \lVert N(t) \rVert - \lVert n \rVert \right) \leq \lVert \lambda \rVert t - \lVert \mu \rVert (t-t_0)(1-\varepsilon(t, K, t_0)).
\]
Since by assumption $\lVert \lambda \rVert < \lVert \mu \rVert$, it is not difficult to choose constants $t, t_0$ and $K$ such that the right hand side is strictly negative (for instance, $t = 1$, $t_0$ small enough and $K$ large enough), which gives the result.

\subsubsection{Proof of Theorem \ref{th:stab2}}

Intuitively, it is clear that the load-dependent RLS strategy performs better than the load-oblivious RLO policy, since it seems harder under RLS to see an empty server. This simple observation shows that the number of empty servers should be part of a Lyapunov function, and indeed this leads us to define the function $f:\mathbb{N}^m\to \mathbb{R}_+$ by:
$$
\forall n,\quad f(n) = \sum_{i=1}^m \max(\epsilon, n_i) = \lVert n \rVert + \varepsilon k_0(n)
$$
with $k_0(n) = \indicator{n_1 = 0} + \cdots + \indicator{n_m = 0}$ the number of empty servers in state $n$. In order for $f$ to be a Lyapunov function, the constant $0<\epsilon<1$ has to satisfy
$$
\epsilon\times \sum_i \mu_i < \sum_i (\mu_i - \lambda_i)-\gamma
$$
for some $\gamma > 0$.

Let $K_0(n)$ (resp.\ $K_1(n)$) be the set of servers that are empty (resp.\ have a single client). Denote by $k_0(n)$ and $k_1(n)$ the respective cardinalities of these sets. Let us compute the average drift $\Delta f(n)$ of the Markov process $N(t)$ under RLS strategy. We have:
\[
\Delta f(n) = \sum_i\lambda_i - \sum_{i\notin K_0(n)}\mu_i +\epsilon \bigg( \sum_{i\in K_1(n)}\mu_i -\sum_{i\in K_0(n)}\lambda_i - Y(n)\bigg),
\]
where $Y(n)$ is the rate in state $n$ at which empty servers are fed by migrating clients. 

\begin{itemize}
\item If $k_0(n)=0$, there is no empty servers in state $n$ and in particular $Y(n)=0$. We have:
$$
\Delta f(n) = \sum_i (\lambda_i-\mu_i)+\epsilon \sum_{i\in K_1(n)} \mu_i < -\gamma,
$$
because of our choice of $\epsilon$.
\item If $k_0(n)>0$, there is at least one empty server in state $n$. Define $p(n)=\max_i n_i$. Considering migrations of the $p(n)$ clients from (one of) the server(s) with maximum size to one of the empty servers, we obtain: $Y(n) \ge {\beta\times p(n)\over m}$, which ensures that $\Delta f(n)<-\gamma$ when $p(n)$ is large enough, say greater than $K$. 
\end{itemize}
We conclude the proof by considering the drift outside the set $F=\{n:f(n)< m(K +\epsilon) \}$. First remark that $F$ is finite. Then, when $n\notin F$, $p(n) \ge K$. We deduce that for all $n\notin F$: $\Delta f(n) <-\gamma$. The positive recurrence follows.

\subsection{Approximate performance estimates}

The system behavior in stationary regime under RLO and RLS strategies is extremely difficult to analyze. For example, $(N(t),t\ge 0)$ is unfortunately not reversible under these strategies. To obtain estimates of the steady state distribution and client sojourn times, we use large-system asymptotics, i.e., we let $m$ grow large. Recently, large-system asymptotics have been successfully applied in many context in communication systems. They have been used for example to understand load balancing issues such as those arising in the supermarket model \cite{Mitz,G00}. In the rest of the section, we denote by $N^{(m)}(t)$ the vector representing the numbers of clients at time $t$ at each server in a system with $m$ servers under either RLO or RLS algorithm.   
 
In what follows, we consider homogeneous systems where $\lambda_i=\lambda$ and $\mu_i=1$ for all $i$. This restriction simplifies the notation and results, but is not essential. We discuss at the end of this subsection how to deal with heterogenous systems. We also assume that the number of clients associated to a given server is bounded by a (possibly very large) constant $B$. Again this assumption is not crucial, and can be relaxed at the expense of a more involved analysis. 

\subsubsection{RLO algorithm}

We first consider RLO algorithms. We assume that a client  jumps from one server to another at the instants of a Poisson process of intensity $\beta$, and that the next server is chosen uniformly at random. The analysis can be extended to any random walk (see \S \ref{subsec:ext}). We represent the system state at time $t$ by $X^{(m)}_k(t)$ the proportion of servers with exactly $k$ clients at time $t$. We also define $S_k^{(m)}(t)=\sum_{l\ge k}X^{(m)}_l(t)$. 

Let us compute the average change in the system state in a small interval of time of duration $dt$, and more specifically the change in $X_k^{(m)}$. Arrivals occur at rate $\lambda m$: An arrival increases $X_k^{(m)}$ if it occurs at servers with $k-1$ clients, and decreases $X_k^{(m)}$ if it occurs at servers with $k$ clients. Hence the change in $X_k^{(m)}$ due to exogenous arrivals is $dt\lambda(X_{k-1}^{(m)}-X_k^{(m)})$. Departures can be analyzed similarly. Let us now compute the change due to client migrations. Clients migrating to server with $k-1$ (resp. $k$) clients increase (resp. decrease) $X_k^{(m)}$. In addition, clients migrating from servers with $k$ (resp. $k+1$) clients decrease (resp. increase) $X_k^{(m)}$. The average change in $X_k^{(m)}$ due to client migrations is thus: $dt\beta((X_{k-1}^{(m)}-X_k^{(m)})\sum_j jX_j^{(m)} - kX_k^{(m)}+(k+1)X_{k+1}^{(m)})$. In summary, the average change in $X_k^{(m)}$ during $dt$ is:
\begin{multline*}
dt\times \bigg[\lambda(X_{k-1}^{(m)}-X_k^{(m)})-(X_k^{(m)}-X_{k+1}^{(m)})\\
+\beta\big[ (X_{k-1}^{(m)}-X_k^{(m)})\sum_j jX_j^{(m)} - kX_k^{(m)}+(k+1)X_{k+1}^{(m)}    \big]\bigg].
\end{multline*}
There is no explicit dependence in $m$, and hence we expect the dynamics of $X_k^{(m)}(t)$ to be close to those of a deterministic solution $x_k$ of the following sets of differential equations: for all $k\in \{0,\ldots,B\}$,
\begin{equation}\label{eq:ode1}
\dot{x}_k =\lambda(x_{k-1}-x_k)-(x_k-x_{k+1}) + \beta \big[(x_{k-1}-x_k)\sum_j jx_j - kx_k+(k+1)x_{k+1}\big],
\end{equation}  
with the convention that $x_{-1}=0=x_{B+1}$. We may write similar differential equations for the evolution of $S_k^{(m)}$. We obtain: for all $k=0,\ldots,B$,
\begin{equation}\label{eq:ode2}
\dot{s}_k =(\lambda+\beta\sum_{j\ge 1}s_j)(s_{k-1}-s_k)-(1+\beta k)(s_k-s_{k+1}),
\end{equation}  
with the convention that $s_{-1}=0=s_{B+1}$. Next we formally justify the above analysis and show that (\ref{eq:ode1}) gives an estimate of system behavior that becomes exact when $m\to\infty$. 

\medskip
\noindent
{\bf Transient regime.} The next theorem states that the approximation is exact over finite time-horizons, and is a direct application of Kurtz's theorem, see Chapter 11 in \cite{EK86}.
\begin{theorem}\label{th:mf1}
Assume that $\lim_{m\to\infty} X^{(m)}(0)=x(0)$ almost surely. Fix $t >0$. We have: almost surely,
\begin{equation}
\lim_{m\to\infty} \sup_{u\le t} \lVert X^{(m)}(u) - x(u)\rVert =0,
\end{equation}
where $x(\cdot)$ is the unique solution of (\ref{eq:ode1}) with initial condition $x(0)$.
\end{theorem}

\bp First, one can easily represent the family of processes $(X^{(m)}(t),t\ge 0)$ as a family of {\it density dependent population processes} as for example defined in \cite{EK86}. Then, define $F:\mathbb{R}^{B+3}\to \mathbb{R}^{B+3}$ by: for all $x\in\mathbb{R}^{B+3}$, $F_{-1}(x)=0=F_{B+1}(x)$ and, for all $k=0,\ldots,B$,
$$
F_k(x)=x_{k-1}(\lambda +\beta \sum_jjx_j)-x_k(\lambda +\beta k+1) +x_{k+1}. 
$$
Now (\ref{eq:ode1}) writes $\dot{x}=F(x)$. $F$ is Lipschitz on ${\mathcal{T}} = \{x\in\mathbb{R}_+^{B+3}: x_{-1}=0=x_{B+1}, \sum_{k=0}^Bx_k=1\}$.  As a consequence, the conditions of the theorem stated in \cite{EK86} p 456 are met, and we deduce the expected result.\ep

\medskip
\noindent
{\bf Stationary regime.} The above theorem holds for finite time-horizons only. It does not say anything about the long-term behavior of the system and in particular for example about the average stationary client sojourn time. To circumvent this difficulty we may use the advanced framework formalized by Sznitman \cite{S91} and further developed in \cite{G00}, and more recently in \cite{nhm09}. Due to space limitations, we skip all details. We invite the reader either to verify that results in \cite{nhm09} apply here or to follow step by step the arguments in \cite{G00} to prove the convergence of the steady-state behavior of finite systems towards the equilibrium point of dynamical system (\ref{eq:ode1}) when $m\to\infty$. More precisely, denote by $X^{(m)}_{eq}$ the stationary empirical distribution of the system with $m$ servers (such distribution exists because $(N^{(m)}(t),t\ge 0)$ is a irreducible finite-state Markov process, and thus positive recurrent). 

\begin{theorem}\label{th:mf2}
Assume that from any initial condition in ${\mathcal{T}}$, the solution of (\ref{eq:ode1}) converges to a unique equilibrium point $\xi$. Then  $X^{(m)}_{eq}$ converges to $\xi$ when $m\to\infty$.
\end{theorem}

From the previous theorem, we know that in a system of $m$ servers, the proportion of servers handling $k$ clients in the stationary regime gets close to $\xi_k$ as $m$ grows large. We may also approximate the average number of clients in the system by $\sum_{k\ge 1} k\xi_k$ and deduce an estimate of the average sojourn time using Little's formula. It remains to show that the system of differential equations (\ref{eq:ode1}) converges to a unique equilibrium point $\xi$, and to characterize $\xi$. 

Let $\xi$ be a fixed point of (\ref{eq:ode1}), then we easily see that: for all $i=1,\ldots,B$,
$$
\xi_i=\xi_0 \times {(\lambda+\beta y)^i\over \prod_{j=1}^i(1+\beta j)},
$$
where $y=\sum_j j\xi_j$. $\xi_0$ is obtained so that $\xi$ is a probability measure. Finally, $y$ must solve:
\begin{equation}\label{eq:fix}
y\times\bigg[1+\sum_{i=1}^B {(\lambda+\beta y)^i\over \prod_{j=1}^i(1+\beta j)}\bigg] = \sum_{i=1}^B i {(\lambda+\beta y)^i\over \prod_{j=1}^i(1+\beta j)}.
\end{equation}
One can check that if $\lambda<1$, (\ref{eq:fix}) indeed has a unique positive solution $y$: if $z=\lambda+\beta y$, $z$ must solve $g(z)=0$ with:
$$
g(z)=(z-\lambda)[1+\sum_{i=1}^B {z^i\over \prod_{j=1}^i(1+\beta j)}] - \sum_{i=1}^B i {z^i\over \prod_{j=1}^i(1+\beta j)}.
$$
The result follows from $g(\lambda)<0$ and $g'(z)\ge 0$ for all $z\ge 0$. In summary the unique equilibrium point of (\ref{eq:ode1}) is $\xi$.

\begin{theorem}\label{th:mfsatb}
From any initial condition $x(0)\in {\mathcal{T}}$, if $\lambda<\mu$, the system of differential equations (\ref{eq:ode1}) converges to the unique equilibrium point $\xi$. 
\end{theorem}

\bp The system enjoys the following important monotonicity property. Consider two initial conditions $x(0)$ and $x'(0)$ such that\footnote{$\le_{st}$ denotes the usual strong stochastic order, i.e., if $x,y$ are probability measures on $\{0,\ldots,m\}$, $x\le_{st}y$ iff for all $j$, $\sum_{i=0}^jx_i\ge \sum_{i=0}^jy_i$.} $x(0)\le_{st} x'(0)$, then if $x$ and $x'$ are the solutions of (\ref{eq:ode1}) with respective initial conditions $x(0)$ and $x'(0)$, we have at any time $t\ge 0$, $x(t)\le_{st}x'(t)$. The proof of this property is based on a probabilistic interpretation of the dynamical system (\ref{eq:ode1}) as the Kolmogorov equations of a collection of birth-death processes of birth rate $\lambda+\beta\sum_jjx_j$ and death rates $(1+\beta k)$ in state $k$. The idea is that for any $s\ge 0$, $x(s)\le_{st}x'(s)$ implies that $\sum_jjx_j(s)\le \sum_jjx_j'(s)$, so the birth rate at time $s$ for $x$ is smaller than that for $x'$, and by a standard coupling argument, we deduce that just after time $s$, we still have $x(s+)\le_{st}x'(s+)$. We may further deduce that this ordering remains valid over time.

Denote by $x^E$ (resp. $x^F$) the solution of (\ref{eq:ode1}) when the system is initially empty $x^E(0)=(1,0,\ldots,0)$ (resp. full $x^F(0)=(0,\ldots,0,1)$). A direct consequence of the above monotonicity property is that $x^E(t)$ (resp. $x^F(t)$) is stochastically increasing (resp. decreasing) over time. For example, for all $h,t\ge 0$, $x^E(t+h)\ge_{st}x^E(t)$. This implies that both $x^E(t)$ and $x^F(t)$ converge to $\xi$ when $t\to\infty$ (since the equilibrium point is unique). We deduce that such convergence also holds starting from any initial condition $x(0)$, since again due to the monotonicity property $x^E(t)\le_{st} x(t)\le_{st}x^F(t)$ for all $t$. \ep

\subsubsection{RLS algorithm} 

The large-system approximation method developed above applies to RLS algorithms. We can similarly derive a deterministic approximation for the evolution of the system empirical measure $X^{(m)}$. When $m\to\infty$, this evolution is characterized by: for all $k=0,\ldots,B$,
\begin{multline}\label{eq:ode3}
\dot{x}_k =\lambda(x_{k-1}-x_k)-(x_k-x_{k+1}) + \\
\beta \left[x_{k-1}\sum_{j\ge k+1}jx_j - x_k\sum_{j\ge k+2}jx_j \quad\quad- kx_k\sum_{j\le k-2}x_j + (k+1)x_{k+1}\sum_{j\le k-1}x_j\right],
\end{multline}  
with by convention $x_{-2}=x_{-1}=x_{B+1}=x_{B+2}=0$. Analyzing the dynamical system (\ref{eq:ode3}) is not straightforward and deserves a full study, which we skip here due to space limitations. In all numerical experiments presented below, we verified the convergence of (\ref{eq:ode3}) to a unique equilibrium point.

\subsubsection{Extension to heterogenous systems and arbitrary random walks (for RLO algorithm)}\label{subsec:ext}

The above asymptotic analysis has been simplified by considering homogenous systems and uniform random walks (for RLO) only. However, in the case of RLS algorithm, it can be easily extended to the case of heterogenous systems, where the arrival rates and server speeds are not identical. To do so, we may classify server according to their arrival rate and speed - servers of the same class have same arrival rate and speed. Then, we can derive a set of differential equations, similar to (\ref{eq:ode1}) or ({\ref{eq:ode3}), approximating the evolution of the proportion of servers of a given class and handling a given number of clients. We obtain a dynamical system whose variables $x_{v,k}$ represent the proportion of servers of class $v$ having $k$ clients. In the case of RLO algorithm, the analysis may also be extended to arbitrary random walks; it suffices to include into the server class the rates at which clients jump towards other servers. For example, servers of class $v$ have the same arrival rate and speed, and the rate at which a client at one of class-$v$ servers jumps to a server of class $v'$ depends on $v$ and $v'$ only. In \cite{nhm09}, the authors present such multi-class asymptotic analysis in details. 

\subsection{Numerical experiments}

We now illustrate the results derived in this section via simple numerical experiments. To evaluate the relative performance of RLO and RLS algorithms, we consider first an homogenous system (for all $i$, $\lambda_i=\lambda$, $\mu_i=1$), and then an extreme heterogenous system where all clients arrive at the same server ($\lambda_1=m\lambda$, and for all $i\ge 2$, $\lambda_i=0$). The system performance is expressed in terms of the average client throughput, defined as the inverse of the average sojourn time. 

\begin{figure}[t]
\centering
\includegraphics[width=0.8\columnwidth]{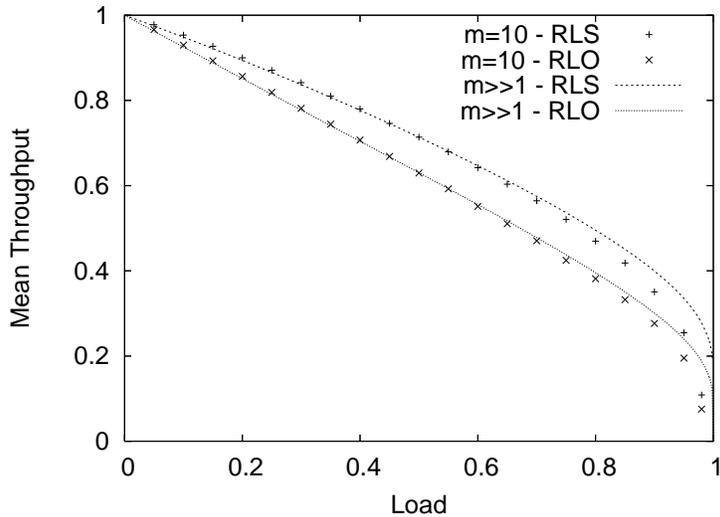}
\caption{Mean throughput under RLS and RLO in homogenous systems as a function of the load $\lambda$. $\beta=0.5$.}
\label{fig1}
\end{figure}

Figure \ref{fig1} gives the average client throughput as a function of $\lambda$ in homogenous systems. We compare the results obtained through the large-system asymptotics $m=\infty$ and those obtained for $m=10$ servers. Note that the asymptotics results are pretty accurate even for small systems. Actually at a load of 0.8, the relative error made in our approximations of the average throughput under RLO and RLS algorithms is less than 4\% when $m=5$, and becomes less than 0.5\% for $m=20$. Note that RLO and RLS are both stable if and only if $\lambda <1$. Surprisingly the performance improvement achieved by the load-dependent RLS algorithm over that obtained under the load-oblivious RLO algorithm is not that significant, typically less than 20\%.

\begin{figure}[t]
\centering
\includegraphics[width=0.8\columnwidth]{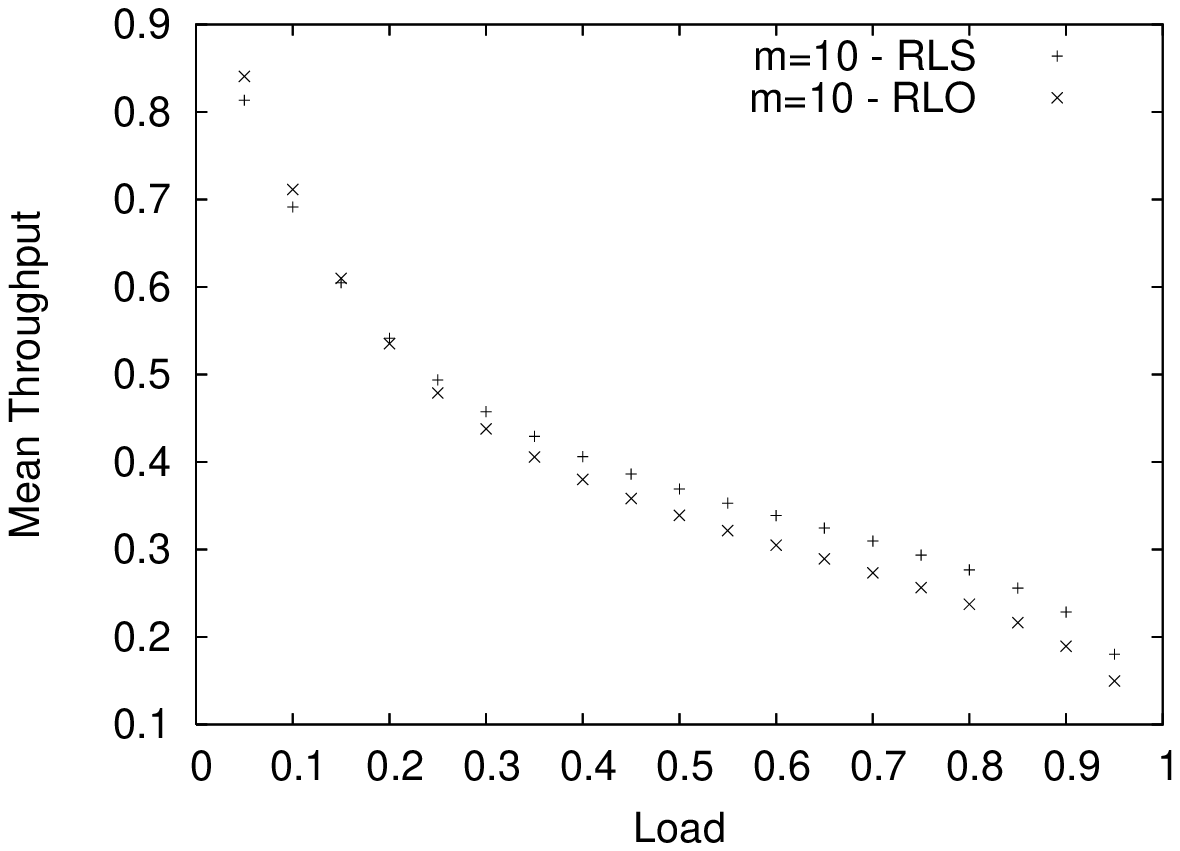}
\includegraphics[width=0.8\columnwidth]{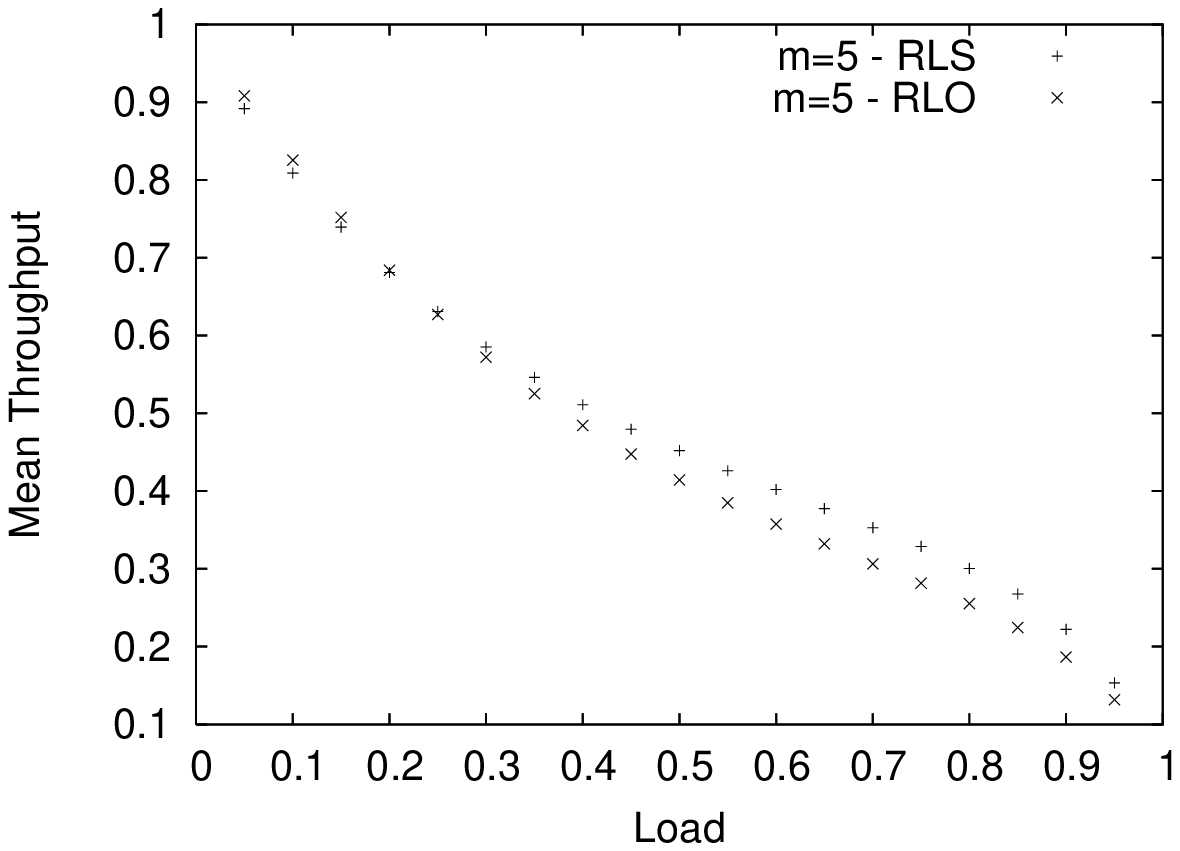}
\caption{Mean throughput under RLS and RLO in heterogenous systems as a function of the load $\lambda$. $\beta=0.5$.}
\label{fig2}
\end{figure}

Figure \ref{fig2} provides the performance in heterogenous systems with $m=5$ and $m=10$. We provide simulation results only, although, as explained above, we could have obtained analytic asymptotic results. Again as expected, even if all clients arrive at the same server, RLS and RLO stabilize the system whenever possible (when $\lambda<1$). The difference between the throughput achieved by RLS and RLO is quite small irrespective of the number of servers considered. Hence it seems that implementing a load-dependent resampling and migration algorithm may not significantly improve the performance.   

\section{Related work}

There have been many studies on distributed, selfish load balancing
algorithms and routing games in closed systems, see e.g. \cite{Koutsoupias99} and references therein. Refer to \cite{nisan} for a quite exhaustive survey. Much of the work in this area has concentrated on finding the fastest sequence of moves that would balance the system, also called Nashification \cite{Feldmann03}. One class of algorithms is the elementary step
system, first described in \cite{Orda93} in which a sequence of best response moves are performed by the clients. Of course this requires that the clients know the status of all the other servers. In \cite{Berenbrink06,Berenbrink07} the authors study closed systems with limited information about the servers' status. They consider a synchronous system where at each step, each server samples a new server randomly and if the load of the sampled server is smaller, then a client moves with probability $(N_c-N_n)/N_c$, where $N_c$ is the load on the current server and $N_n$ is the load of the sampled server. It is shown that the expected time to balance the system is $O(\log \log m + n^4).$ A modification of this load balancing algorithm is studied in \cite{Berenbrink07}, and it is shown that  the expected time to balance the system is $O(\log m + n \log n)$. In \cite{Goldberg04}, the author considers clients dynamics identical to those considered in this paper and uses the potential function introduced in \cite{Even-Dar07} to quantify the time to achieve system balance. It is shown that the expected time to reach a balance scales at most as  $O(m^2).$ We provide significant improvements on this bound.

In open systems, the client moves interact in a complicated manner with the client arrival and departure processes. There is very little work trying to understand this interaction. None of the existing work deals with a system similar to that studied here. For instance, \cite{AK08} analyzes the interaction in a game-theoretical framework, where arrivals are {\it adverserial}, and where a central controller moves clients with the aim of stabilizing the system. The performance of the classical {\it work stealing} load-balancing scheme has also been studied, see e.g. \cite{BF03} and references therein. Of course there is an abundant literature on the performance of classical load-balancing schemes in open systems where clients are assigned to a given server for the entire duration of their service, see e.g. the analysis of the supermarket model in \cite{Mitz,G00}. To our knowledge, the present paper provides the first analysis of natural distributed resampling and migration strategies in open systems.

\section{Conclusion}

In this paper, we have analyzed the performance of distributed load balancing schemes where clients independently decide to resample and change server to improve their service rate. We considered two natural random resampling and migration strategies: A load oblivious strategy RLO where clients randomly move from one server to another without accounting for the actual server loads, and a load-dependent selfish strategy RLS where clients  randomly resample servers and migrate only if their rate is improved. 

In closed systems where the population of clients is fixed, we have provided a new tight bound on the time to balance server loads under RLS strategy. This time can be interpreted as the time to reach a Nash Equilibrium in this selfish routing game. Our bound considerably improves the bounds available in the literature. But it holds only in the case of homogenous systems where servers have identical service rates. It seems challenging and interesting to figure out how to apply our methodology to obtain bounds on the time to balance the system in the case of heterogenous systems. It might also be interesting to investigate the time it takes to balance the system in scenarios where client migrations are limited, in the sense that from a given server, clients can migrate to a restricted subset of servers (as for example specified via a graph).

In open systems where clients arrive at the various servers at different rates, we provided a first analysis of the system dynamics. These dynamics are complicated as the client arrival and departure processes interact with the client migration processes. We have shown that both RLO and RLS load balancing strategies are able to stabilize the system whenever this is at all possible. It may appear somehow surprising that a completely distributed and load-oblivious algorithm such as RLO can achieve maximum stability. Using large-system asymptotics, we also provided approximate estimates of the mean client sojourn time. The results show that again, surprisingly, the load-oblivious RLO strategy does not yield significant performance losses compared to the load-dependent RLS strategy. These findings are valid for exponential service requirements, and it would be interesting to know whether they remain valid for other service requirement statistics. 

An interesting extension of the present work (especially relevant when considering spectrum sharing issues) is to analyze the case where clients may use resources from several servers simultaneously. There are some preliminary results in this direction in \cite{MassKeyTow}, but neither the time to reach equilibrium or the population dynamics are studied.

\end{document}